\documentclass{ieeetj}
\usepackage{cite}
\usepackage{amsmath,amssymb,amsfonts}
\usepackage{algorithmic}
\usepackage{multirow}
\usepackage{graphicx,color}
\usepackage{textcomp}
\usepackage{xcolor}
\usepackage{hyperref}
\usepackage{algorithm,algorithmic}
\usepackage{booktabs}
\usepackage{subfigure}
\usepackage{mathtools}
\usepackage[square,numbers]{natbib}
\def\BibTeX{{\rm B\kern-.05em{\sc i\kern-.025em b}\kern-.08em
    T\kern-.1667em\lower.7ex\hbox{E}\kern-.125emX}}
\AtBeginDocument{\definecolor{tmlcncolor}{cmyk}{0.93,0.59,0.15,0.02}\definecolor{NavyBlue}{RGB}{0,86,125}}

\def\OJlogo{\vspace{-4pt}$<$Society logo(s) and publication title will appear here.$>$}
\def\seclogo{\vspace{10pt}$<$Society logo(s) and publication title will appear here.$>$}

\def\authorrefmark#1{\ensuremath{^{\textbf{#1}}}}

\begin{document}
\receiveddate{XX Month, XXXX}
\reviseddate{XX Month, XXXX}
\accepteddate{XX Month, XXXX}
\publisheddate{XX Month, XXXX}
\currentdate{XX Month, XXXX}
\doiinfo{XXXX.2022.1234567}

\markboth{}{Author {et al.}}

\title{Adversarial Representation Learning for Robust Privacy Preservation in Audio}

\author{Shayan Gharib\authorrefmark{1*}, Minh Tran\authorrefmark{2*}, Graduate Student Member,  Diep Luong\authorrefmark{2}, Konstantinos Drossos\authorrefmark{3{**}}, Member, IEEE, and Tuomas Virtanen\authorrefmark{2}, Fellow, IEEE}
\affil{Department of Computer Science, University of Helsinki, 00014 Helsinki, Finland}
\affil{Faculty of Information Technology and Communication Sciences, Tampere University, 33100 Tampere, Finland}
\affil{Nokia Tech, 02610 Espoo, Finland}
% \corresp{Corresponding author: First A. Author (email: author@ boulder.nist.gov).}
% \authornote{This paragraph of the first footnote will contain support information, including sponsor and financial support acknowledgment. For example, ``This work was supported in part by the U.S. Department of Commerce under Grant 123456.''}

\begin{abstract}
Sound event detection systems are widely used in various applications such as surveillance and environmental monitoring where data is automatically collected, processed, and sent to a cloud for sound recognition. However, this process may inadvertently reveal sensitive information about users or their surroundings, hence raising privacy concerns.
In this study, we propose a novel adversarial training method for learning representations of audio recordings that effectively prevents the detection of speech activity from the latent features of the recordings. The proposed method trains a model to generate invariant latent representations of speech-containing audio recordings that cannot be distinguished from non-speech recordings by a speech classifier. The novelty of our work is in the optimization algorithm, where the speech classifier's weights are regularly replaced with the weights of classifiers trained in a supervised manner. This increases the discrimination power of the speech classifier constantly during the adversarial training, motivating the model to generate latent representations in which speech is not distinguishable, even using new speech classifiers trained outside the adversarial training loop. The proposed method is evaluated against a baseline approach with no privacy measures and a prior adversarial training method, demonstrating a significant reduction in privacy violations compared to the baseline approach. Additionally, we show that the prior adversarial method is practically ineffective for this purpose.
\end{abstract}


%\IEEEspecialpapernotice{(Invited Paper)}

\maketitle
\def\thefootnote{*}\footnotetext{Equally contributing authors}\
\def\thefootnote{**}\footnotetext{Contribution during affiliation with Tampere University}

\section{Introduction}
\label{submission}

% The availability of data has never been easier using the emergence of ever-present devices that employ sensors to ceaselessly collect and process our data leading to the state-of-the-art performances in many machine learning tasks \cite{chung2021w2v, NEURIPS2020_1457c0d6, yu2022coca}.
% The increasing availability of data has been facilitated by the proliferation of ever-present devices equipped with sensors that continuously collect and process large amount of data, enabling remarkable advancements in many machine learning tasks \cite{chung2021w2v, NEURIPS2020_1457c0d6, yu2022coca}.
The proliferation of ever-present devices equipped with sensors has led to an exponential increase in data availability. These devices continuously collect and process large amounts of data, facilitating remarkable advancements in various machine learning tasks \cite{chung2021w2v, NEURIPS2020_1457c0d6, yu2022coca}.
However, this trend raises concerns regarding the privacy of users' personal information both during the data collection process and when the data is utilized by machine learning models \cite{kumar15, glackin17}. Speech interfaces and acoustic monitoring are prominent areas among those with active research focusing on preserving user data privacy. These systems record audio which may contain biometric information such as human voices that can be identified and attributed to individuals. Therefore new legislation has been enacted to safeguard users against the inherent risks associated with the exposure of personal information \cite{pererocodosero22}.
% One of the applications with highly active research regarding the privacy preservation of users' data is in speech interfaces and acoustic monitoring.
% Therefore, new legislation has been adopted to protect users against this inherent vulnerability of personal information \cite{pererocodosero22}.
\begin{figure}[t]
    \begin{center}
    \centerline{\includegraphics[width=\columnwidth]{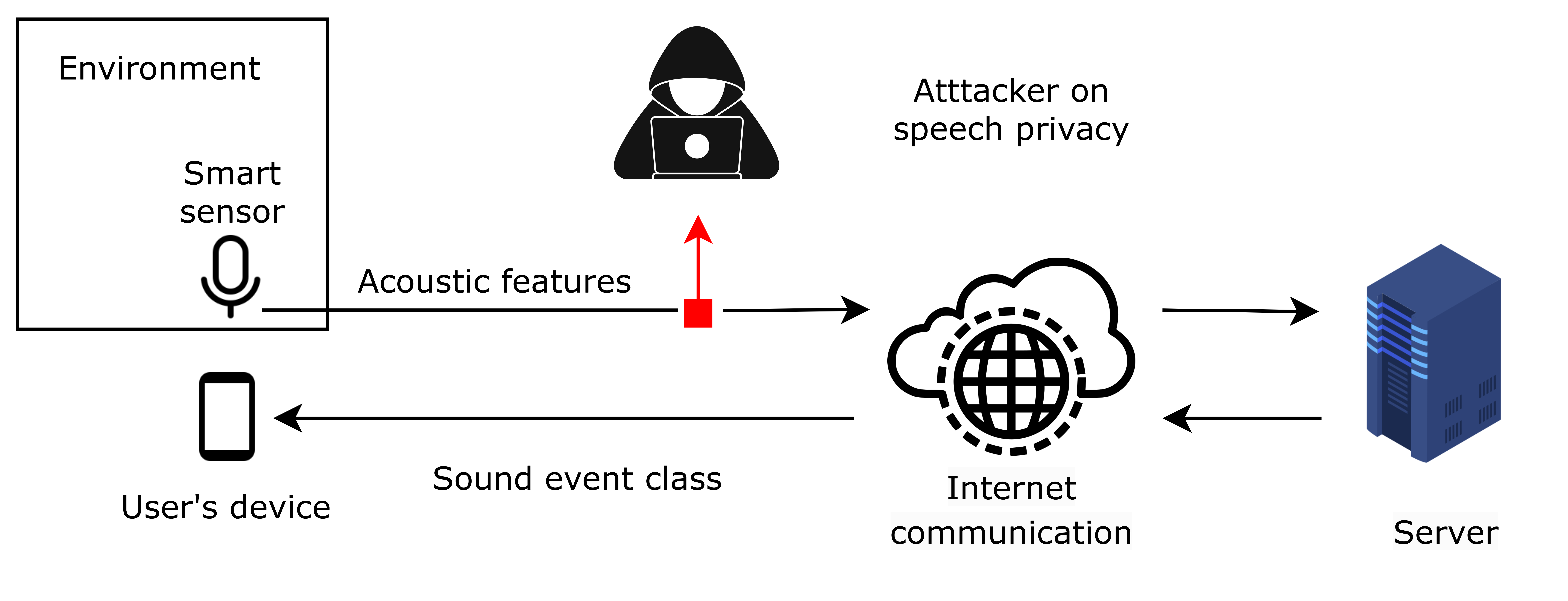}}
    \caption{Illustration of the problem setup where speech privacy is compromised during the transmission of acoustic features to a cloud platform.}
    \label{problem-setup-diagram}
    \end{center}
\end{figure}
Acoustic  pattern classification has numerous applications in smart cities, smart homes, and context-aware devices \cite{Krstulovic2018}. These systems aim to automatically detect targeted sound events such as sirens, birds chirp, and window breakage, among others. While the primary focus of these methods may not be on speech-related information, the environments where they operate often include speech.
% They are typically targeting to automatically recognize sound events relevant to their applications, such as sirens, gunshots, window breakage, etc. 
% Even though these methods may not be targeting information in speech, the environments where these methods are used often contain speech as well. 
Human speech contains a significant amount of personal information, including speakers' identity, gender, accent, or sensitive content discussed during conversations \citep{pererocodosero22, williams21}. Many voice interface and acoustic monitoring systems, including daily consumer devices such as smartphones, locally extract features from audio and transmit them to a cloud for recognition tasks \cite{hossain19}. Unauthorized access to this information by adversaries can have detrimental consequences for the individuals involved. Figure~\ref{problem-setup-diagram} provides an illustration of a typical setup for this scenario, highlighting that the disclosure of such information should only occur with speakers' consent.
% If adversaries get access to any of this information, the effect might be detrimental to the speaker. A typical setup for this use case has been illustrated in Figure~\ref{problem-setup-diagram}. Thus, this information can only be exposed with the speaker's agreement.

An example of this scenario, which served as the primary motivation for conducting this research, pertains to automatic sound recognition devices employed in home care settings. 
% which also was the initial motivation of this research, is automatic sound recognition devices for home care use. 
The objective was to develop a device capable of promptly notifying nurses when an elderly individual is in a dangerous situation requiring assistance. The system actively monitors the surrounding soundscape to trigger an alert in the event of an emergency. Therefore, it inevitably captures many speech signals. To ensure privacy, it becomes imperative for recording devices to hide the speech-related information within the encoded features of the signals to protect such information during the data transmission.

To address this challenge, the objective of this study is to integrate a privacy-preserving algorithm into the representation learning process in order to base final classification decisions on features that ensure users' privacy. To achieve this, we employ an adversarial learning setup based on deep neural networks (DNNs) to create latent representations of audio recordings that contain information required for the recognition of targeted sound events, while removing information that could be used for speech analysis. 

Inspired by \citet{pmlr-v37-ganin15}, our adversarial setup includes two neural networks: a feature extractor and a speech classifier. The feature extractor is designed to manipulate the latent features in such a way that it confuses the speech classifier, thereby reducing its performance on speech classification tasks. 

Although the general idea has been used previously in a different application, this approach is vulnerable to the retrieval of speech attributes \cite{srivastava2019}, especially when the learned latent features are used to re-train a separate speech classifier outside the adversarial process. To address this vulnerability, we propose a straightforward solution that can be seamlessly integrated into the learning process. The speech classifier is frequently replaced by a new one, that is trained until convergence, outside of the adversarial learning process. This approach ensures that the speech classifier is not easily tricked by the feature extractor, creating a robust training process that ensures the speech-related information is not retrievable. Throughout this paper, we refer to this algorithm as \textit{robust discriminative adversarial learning} (RDAL).
% The key technical factor of our method is a simple solution that can be easily integrated into the learning process, where the speech classifier is frequently trained until convergence outside the adversarial learning process, to avoid being tricked easily by the feature extractor. Therefore, it creates a robust training process that ensures the speech information is not retrievable. We refer to this algorithm throughout the paper as \textit{robust discriminative adversarial learning (RDAL)}.   

% Therefore, the contribution of this paper is the proposal of the RDAL algorithm as the first sound event classification system that successfully prevents the detection of speech in acoustic data. This demonstrates the effective application of RDAL for privacy-preserving sound event classification and potentially for other machine listening tasks. Furthermore, we introduce a publicly available dataset as the first benchmark dataset for privacy-preserving in sound event classification tasks; to our best knowledge, such a dataset has not been available beforehand. 

% The main contribution of this paper is the introduction of the RDAL algorithm, which create an adversarial learning representation that effectively prevents the detection of speech in learned representations of acoustic features, enabling robust sound event classification.
% serves as the first sound event classification system to effectively prevent the detection of speech in acoustic data. 
The main contribution of this paper lies in the introduction of the RDAL algorithm, which facilitates the generation of adversarial learning representations that effectively prevent the detection of sensitive information within the latent feature space derived from acoustic features, thereby enabling robust sound event classification. 
% A preliminary version of this work was previously published as a conference paper \cite{luong2023representation}. The current version considerably extends the the work of \citet{luong2023representation}  by offering comprehensive insights into the RDAL algorithm. It provides detailed explanations about the algorithm itself, the rationale behind its usage, and its effective approach in mitigating vulnerabilities observed in prior studies. Additionally, we expand our dataset from our previous work in \cite{luong2023representation} to enhance the understanding of RDAL's generalizability across various sound event classes. 
A preliminary version of this work was previously published as a conference paper \cite{luong2023representation}. This paper introduces the RDAL algorithm, which was not previously covered in the conference paper.
% The current version significantly expands upon our recent work by providing comprehensive insights into the RDAL algorithm. 
While the conference paper primarily focused on evaluating the effectiveness of source separation in conjunction with RDAL, it did not delve into the detailed explanations, rationale, and vulnerabilities addressed by RDAL. In this extended version, we present comprehensive insights into the RDAL algorithm, including a thorough exploration of the algorithm itself, its underlying principles, and its efficacy in mitigating observed vulnerabilities in prior studies. Furthermore, to enhance the understanding of RDAL's generalizability across various sound event classes, we expand our dataset from our previous work in \cite{luong2023representation}. Lastly, we incorporate gender classification into our evaluation setup to demonstrate RDAL's performance on this novel task.

\section{Related work}

% In this section, we review prior works on privacy preservation in two audio related tasks: Machine listening, speech recognition. Although, in machine listening tasks, learning privacy preserving features in an adversarial training setup has not been studied yet, there are few attempts that successfully employ adversarial training for anonymizing speech characteristics such as speaker identity in automatic speech recognition systems. 

In this section, we discuss previous works on privacy preservation in two audio-related tasks: machine listening and speech recognition. While the application of adversarial training for learning privacy-preserving features in machine listening tasks has not been extensively explored, there have been successful attempts to use adversarial training to anonymize speech characteristics, such as speaker identity, in automatic speech recognition (ASR) systems.

\subsection{Privacy-preservation in machine listening tasks}

% \citet{larson11} aim to create a cough detection system on mobile phones that enable higher level features of audio invertible to original sounds in order to be evaluated by physicians. The privacy scope defined in this paper is to make speech unintelligible in the reconstructed audio. The paper suggests using principal component analysis (PCA) analysis on the spectrogram of the cough sound to create a set of selected eigenvectors with the biggest eigenvalues. The reconstruction of audio using these features is expected to make speech unintelligible while retaining enough information for cough detection. In spite of that, the limitation in learning capacity of PCA reduces the generalizability performance of such systems especially when the task become polyphonic sound event detection of different events. In addition, unintelligible speech in the recovered audio signal does not guarantee that speech information cannot be extracted either. 
\citet{larson11} developed a cough detection system for mobile phones, aiming to render speech unintelligible in the reconstructed audio. They used principal component analysis (PCA) analysis on cough sound spectrograms to select eigenvectors with the biggest eigenvalues for audio reconstruction. However, PCA's limited learning capacity reduced system performance, especially for polyphonic sound event detection. Additionally, unintelligible speech in the recovered audio does not guarantee that speech information cannot be extracted.
% \citet{wang19} state that human speech largely belongs to the range between 80~Hz and 3 kHz, and propose a bandstop filter to filter out speech in the audio signal prior to the recognition of indoor human activity. Although speech is filtered out, filtering will also lead to the loss of information of other sound events and negatively affect the machine listening performance, especially for the events that have a largely overlapping frequency range with human speech.
\citet{wang19} assert that human speech predominantly falls within the frequency range of 80~Hz to 3 kHz. To recognize indoor human activity, they suggest using a bandstop filter to remove speech from the audio signal. However, this filtering process also results in the loss of information related to other sound events. Consequently, machine listening performance is adversely affected, particularly for events that share a significant frequency range with human speech.

% \citet{nelus21} propose a privacy-preserving representation learning based on variational information extracted using DNNs. The objective is to generate latent representations from the low-level audio features, i.e. log mel-band energy, that carry maximum information for the classification task while the speaker information is minimized. They do so by minimizing the mutual information between mel-band energy and extracted features by DNNs, indicating there is little dependency between these two features. Simultaneously, the model is trained to make sure the extracted latent features are informative enough for the classification task.
\citet{nelus21} propose privacy-preserving representation learning using DNNs to extract variational information. Their objective is to generate informative latent representations from log mel-band energy for the classification task while minimizing speaker information. This is achieved by minimizing the mutual information between mel-band energy and extracted features by DNNs, ensuring low dependency between these two features. Simultaneously, the model is trained to keep the extracted latent features informative for classification.

\subsection{Adversarial privacy-preserving representations in speech recognition tasks}
% Initially, related research in speech recognition tasks, used adversarial representation learning as a tool to induce robustness to the classification performance of predictive models by ignoring irrelevant information. In this regard, following works utilize adversarial learning to create speaker-invariant features, since speaker variability can negatively affect the performance of acoustic modelling systems. 
In previous research on speech recognition tasks, adversarial representation learning was employed to enhance the robustness of predictive models by disregarding irrelevant information. Specifically, subsequent studies utilized adversarial learning to generate speaker-invariant features, as speaker variability can have a detrimental impact on the performance of acoustic modeling systems.

% Inspired by \citet{pmlr-v37-ganin15}, \citet{meng18} minimize speaker information in the audio representation for senone classification task by creating a minimax speaker classification objective to produce bottleneck features that are speaker-invariant, yet discriminative for senone predictions. Similarly, \citet{tsuchiya18} developed a speaker-invariant representation of audio data for acoustic modelling of zero-resource language by adversarial training of a feature extractor network. Such studies are comparable to privacy-preserving representation learning in that information about speakers is eliminated from the extracted features.
% Inspired by \citet{pmlr-v37-ganin15}, 
\citet{meng18} minimize speaker information in audio representations for senone classification. They introduce a minimax speaker classification objective to generate bottleneck features that are speaker-invariant and yet discriminative for senone predictions. Similarly, \citet{tsuchiya18} employ an adversarial training setup to develop a speaker-invariant representation of audio data for zero-resource language acoustic modeling. These studies share similarities with privacy-preserving representation learning as they eliminate speaker information from the extracted features.

% One of the first studies of privacy preservation in audio domain using adversarial learning has been done by \citet{srivastava2019}. They aim to hide speaker identities in ASR task by anonymizing the latent representations of an end-to-end ASR network. This is achieved by a minimax objective between encoded representations obtained by a speech encoder and a speaker classifier. Although the system can reduce the performance on close-set speaker identification, some residual information about speakers can still be recovered, resulting in an increase in the performance of open-set speaker verification.
\citet{srivastava2019} conducted one of the initial studies on privacy preservation in the audio domain using adversarial learning. Their objective was to conceal speaker identities in ASR systems by anonymizing the latent representations of an end-to-end ASR network. This was achieved through a minimax objective between the encoded representations from a speech encoder and a speaker classifier. While the system reduced performance in close-set speaker identification, residual information about speakers could still be recovered, leading to improved performance in open-set speaker verification.

% Instead of removing general information about speakers' identity, later works would investigate more into removing certain speaker attributes. \citet{noe21} preserve the privacy in speakers' gender information in an automatic speaker verification system using similar adversarial training. While this study only investigates one attribute of speakers, in reality, there may be the need for hiding several types of information. \citet{pererocodosero22} reconstruct privacy-preserving x-vectors with multiple adversarial privacy domains related to different speaker attributes: ID, gender, and accent. They also show that the evaluation of utility tasks as well as the privacy performance improve with multiple adversarial privacy domains. However, the recoverability of sensitive speech information in audio features after adversarial learning in \citet{srivastava2019} has yet to be addressed by any proposed adversarial learning systems. The privacy-preserving features obtained by feature extractor networks inside the adversarial training loop do not guarantee that the sensitive information is completely removed and cannot be recovered.
Later studies focused on removing specific speaker attributes instead of general speaker identity information. For example, \citet{noe21} used adversarial training to preserve the privacy of speakers' gender information in an automatic speaker verification system. While this study investigated a single attribute, in practice, there may be a need to conceal multiple types of information. \citet{pererocodosero22} reconstructed privacy-preserving x-vectors using multiple adversarial privacy domains related to different speaker attributes, such as ID, gender, and accent. They demonstrated that incorporating multiple adversarial privacy domains improved both utility tasks and privacy performance. However, the recoverability of sensitive speech information in audio features after adversarial learning, as observed in \cite{srivastava2019}, has not been addressed by proposed adversarial learning systems. The privacy-preserving features obtained through adversarial training do not guarantee complete removal of sensitive information or prevention of recovery.

\section{Method}
\label{method}

\subsection{Problem setup} 

% We consider a scenario where the aim is to identify a utility attribute $\mathbf{y}$ using a latent representation $\mathbf{z}$ calculated from audio features $\mathbf{x}$ while making sure that $\mathbf{z}$ contains minimal information regarding a sensitive attribute $\mathbf{s}$. For the purpose of this paper, we consider specific targeted sound event classes as our utility $\mathbf{y}$ and speech presence as our sensitive attribute $\mathbf{s}$, however, these could be trivially extended to any other attributes. For instance, the sensitive attribute can be simply replaced by speaker identity, accent, gender, etc. In this study, we limit ourselves using speech presence estimation only, since that is the most prominent type of speech information. We assume that a method that is able to remove information about speech presence would also be able to remove information about other speech characteristics, since those are in general more difficult to recognize than speech presence.
In this study, we focus on identifying a utility attribute $\mathbf{y}$ using a latent representation $\mathbf{z}$ derived from acoustic features $\mathbf{x}$. Our goal is to ensure that $\mathbf{z}$ contains minimal information related to a sensitive attribute $\mathbf{s}$. While our specific utility attribute is targeted sound event classes and the sensitive attribute is speech presence, our approach can be extended to other attributes, such as speaker identity, accent, or gender. Given that speech presence is a prominent type of speech information, our focus is primarily on speech presence estimation. We assume that a method capable of removing information about speech presence would also be effective in removing information about other speech characteristics, which are generally more challenging to recognize.
Following the above problem setup, we assume that we have access to a labeled dataset $\mathbb{X}$, consisting of $N$ data samples $\mathbf{x}$ accompanied by sound event labels $\mathbf{y}$ and speech labels $\mathbf{s}$, i.e. $ \mathbb{X} = \{(\mathbf{x}_{i}, \mathbf{y}_{i}, \mathbf{s}_{i})\}_{i=1}^{N}$.    

\begin{figure*}[t!]
    \begin{center}
    \centering
    \includegraphics[width=0.7\textwidth]{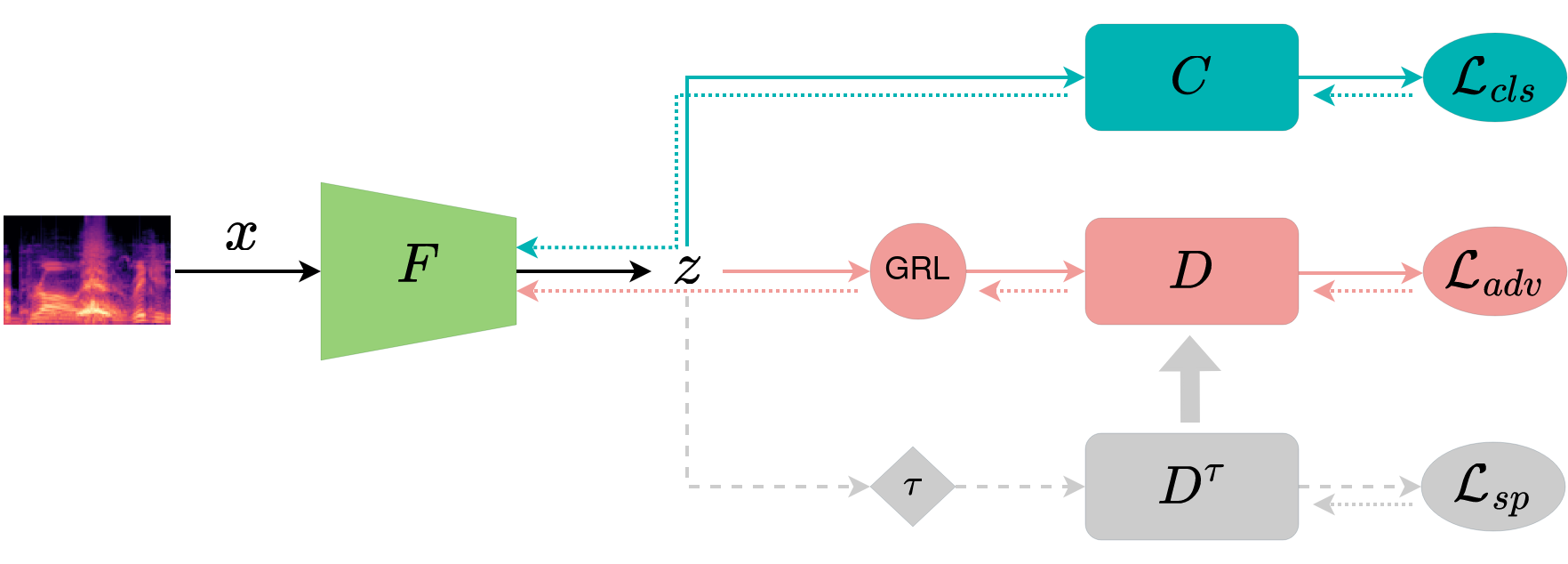}
    \caption{Schematic diagram of the proposed method. $F$, $C$, $D$, and $D^\tau$ are neural networks and $\mathcal{L}$ denotes different loss terms employed in our method. The solid lines illustrate the regular forward pass. The dashed line actives after $\tau$ epochs. Finally, the dotted lines represent the backpropagation of each specific error w.r.t the associated parameters.}
    \label{system-architecture-figure}
    \end{center}
\end{figure*}

Given an input $\mathbf{x}$, our goal is to compute a latent representation $\mathbf{z}$ using a feature extractor $F$, i.e. $ \mathbf{z} = F(\mathbf{x})$, that enables a classifier $C$ to perform multi-class classification for targeted sound events where the goal is to classify each input into one of the predefined sound event classes denoted as $\mathbf{y} \in \{1, 2, ..., Y\}$, resulting in an estimated class $\hat{\mathbf{y}} = C(\mathbf{z})$. To prevent disclosing speech-related information, the latent representation $\mathbf{z}$ should not reveal any indications that allow classification of $\mathbf{x}$ into its speech class $\mathbf{s}$ using a speech classifier $D$, i.e. $\mathbf{\hat{s}} = D(\mathbf{z})$. We formulate this problem such that both goals are met simultaneously. Figure~\ref{system-architecture-figure} illustrates each component of our method and their interconnection.

% Figure~\ref{system-architecture-figure} shows each component of our method and how they are connected to each other.

\subsection{Robust discriminative adversarial learning}
We build upon a discriminative adversarial learning approach which was initially introduced by \citet{pmlr-v37-ganin15} for the purpose of obtaining domain invariant representations in an unsupervised domain adaptation task.

To achieve a well performing sound event classification, the feature extractor $F$ and the sound event classifier $C$ are jointly trained to predict the present sound event in an input, i.e. $\hat{\mathbf{y}}_{i} = C(F(\mathbf{x}_{i}))$. As the first part of our algorithm, the objective function
\begin{equation}
\label{eq:l_cls}
\min_{F, C}\mathcal{L}_{cls} = -\mathbb{E}_{(\mathbf{x}, \mathbf{y}) \sim \mathbb{X}} \sum_{i=1}^{N} \mathbf{1}_{[i=\mathbf{y}]} \log(\hat{\mathbf{y}}_{i}),
\end{equation}
is then minimized by optimizing the parameters of $F$ and $C$ in order to reduce the classification error between the true labels $\mathbf{y}_i$ of targeted sound events and their corresponding predictions $\hat{\mathbf{y}}_i$.

In order to prevent the recognition of any speech information, we use an adversarial training method consisting of two components: the feature extractor $F$ and the speech classifier $D$. More specifically, $D$ is employed to predict present speech in an input $\mathbf{x}_i$ based on latent features $\mathbf{z}_i$. This is achieved using the objective function
\begin{equation}
\label{eq:L_adv}
\max_{F}\min_{D} \mathcal{L}_{adv} = \mathbb{E}_{(\mathbf{x}, \mathbf{s}) \sim \mathbb{X}} \sum_{i=1}^{N} \ell(\mathbf{s}_i, D(F(\mathbf{x}_i))).
\end{equation}
We use binary cross entropy as the loss function $\ell$ in our experiments.
% The aim of $D$ is to maximize its performance on the classification of speech. However, the feature extractor $F$, as a component in the adversarial learning algorithm, tries to confuse $D$ in order to minimize its ability for classifying speech from the internal representations $\mathbf{z}_{i}$. This is achieved using a gradient reversal layer (GRL) module which is placed between $D$ and $F$ connecting these two networks to each other \cite{pmlr-v37-ganin15}. GRL functions differently in forward and backward propagation of information through networks. Its functionality resembles an identity matrix in forward propagation, hence the output of this layer is the same as its input. However, in the backward propagation, it multiplies the partial derivatives of the $\mathcal{L}_{adv}$ loss with respect to the parameters of the feature extractor $\theta_{F}$, i.e. $\frac{\partial\mathcal{L}_{adv}}{\partial\theta_{F}}$, with $-\lambda$, where $\lambda \geq 0$. This means that when $\lambda=0$, the parameters of $F$ are optimized only using the objective $\mathcal{L}_{cls}$, and as $\lambda$ increases the contribution of $\mathcal{L}_{adv}$ in optimizing $\theta_F$ improves as well. As a result, this encourages $F$ to produce invariant representations of its inputs so that the information about the speech is discarded. Consequently, this undermines the performance of $D$ in classifying speech classes. 
The primary objective of the speech classifier $D$ is to achieve optimal performance in speech classification. However, within the adversarial learning framework, the feature extractor $F$ aims to obfuscate $D$ by minimizing its ability to classify speech based on the internal representations $\mathbf{z}_{i}$. To facilitate this, a gradient reversal layer (GRL) module is introduced, connecting $D$ and $F$ in the network \cite{pmlr-v37-ganin15}. The GRL operates differently during forward and backward propagation. In the forward pass, it functions as an identity mapping, preserving the input as the output. However, during backward propagation, it multiplies the partial derivatives of the adversarial loss $\mathcal{L}_{adv}$ with respect to the feature extractor parameters $\theta_F$, denoted as $\frac{\partial\mathcal{L}_{adv}}{\partial\theta_{F}}$, by a negative coefficient $-\lambda$, where $\lambda \geq 0$. This implies that when $\lambda=0$, the feature extractor $F$ is solely optimized based on the classification objective $\mathcal{L}_{cls}$. As $\lambda$ increases, the contribution of $\mathcal{L}_{adv}$ in optimizing $\theta_F$ becomes more pronounced. Consequently, $F$ is encouraged to generate invariant representations that discard speech-related information, thereby undermining the speech classification performance of $D$. Therefore, we can summarize the optimization process of $F$ as
\begin{equation}
\label{eq:param_update}
\theta_F \longleftarrow \theta_F - \mu \left(\frac{\partial \mathcal{L}_{cls}}{\partial \theta_F} - \lambda \frac{\partial \mathcal{L}_{adv}}{\partial \theta_F}\right)
\end{equation}
where $\mu$ is the learning rate.

% Despite reaching the convergence of the minimax objective in Equation~\ref{eq:L_adv}, and in return, reducing the performance of speech classification task for speech classifier $D$ during the adversarial training, this method does not guarantee that the generated representations $\mathbf{z}$ are deprived of information regarding the sensitive attribute $\mathbf{s}$. This can be easily validated by simply training a new speech classifier over $\mathbf{z}$ outside the adversarial learning process. This problem has previously been shown in \cite{srivastava2019} for anonymization of speakers in an ASR system. \citet{srivastava2019} concluded that the method suffers from generalization performance due to the limitation in the representation capacity of the adversarial branch. In addition, this problem is not limited to privacy-preserving representations learned in an adversarial setting with a minimax objective. 
Although the minimax objective in Equation~\ref{eq:L_adv} leads to the convergence of the adversarial training and subsequently reduces the speech classification performance of the speech classifier $D$, it does not guarantee that the resulting representations $\mathbf{z}$ are free from sensitive attribute information $\mathbf{s}$. This limitation becomes evident when training a new speech classifier solely on $\mathbf{z}$ outside the adversarial learning process. A similar issue was identified by \citet{srivastava2019}, who aimed to anonymize speakers in an ASR system. \citet{srivastava2019} found that the method's generalization performance was hindered by limitations in the representation capacity of the adversarial branch. It is important to note that this problem extends beyond privacy-preserving representations learned within an adversarial framework.

In a related study conducted by \citet{Jin_2021_ICCV}, they address the same issue from a broader perspective by aiming to minimize distribution shift in the latent space through a discriminative adversarial setup, similar to our approach. As the adversarial training progresses, the alignment between the distributions of latent features for different speech classes increases, resulting in reduced discriminative ability of the speech classifier $D$ in distinguishing between them. Consequently, the feature extractor $F$ has less incentive to further align the latent representations, posing challenges for optimizing $F$ and $D$ to learn invariant representations within this adversarial setting. 
% dist of probabilities of predictions vs dist. of latent features
% The maximum uncertainty is achieved when the $p=\frac{1}{2}$, and this in turn means maximum predictions uncertainty for the discriminator $D$.  
% leads to an increased uncertainty of the speech classifier predictions. The uncertainty in predictions of speech classifier can be quantified using the higher entropy of speech classification, leading to a reduced discriminative ability of the discriminator in distinguishing between the two classes. The alignment can be quantified using the entropy of binary classification
% In our work, we tackle the same challenge where our speech classifier $D$ determines the presence of speech in data samples. As the distributions of speech and non-speech samples align more closely, the uncertainty of model predictions increases. 

To address such vulnerability in the optimization of adversarial branch, we propose a mechanism aimed at enhancing the discrimination power of the speech classifier $D$ and ensuring the generation of generalizable and robust privacy-preserving latent features. RDAL introduces a supervised training step using the latent representations $\mathbf{z}$ after every $\tau$ epochs of adversarial training to train a new speech classifier, denoted as $D^\tau$, in a supervised manner. Subsequently, the parameters of $D$ are updated using those of $D^\tau$ before continuing adversarial process. This iterative process compels the feature extractor $F$ to continuously modify its outputs, making the representations of speech classes indistinguishable so that the new experts/speech classifiers are not able to distinguish between them. This process is repeated until further training iterations no longer lead to improved performance of $D^\tau$. In our study, speech labels can be represented using binary labels $\mathbf{s}$, and the training of $D^\tau$ is done by minimizing 
\begin{multline}
\label{eq:verify_loss}
\min_{D^\tau}\mathcal{L}_{sp} = -\mathbb{E}_{(\mathbf{x}, \mathbf{s}) \sim \mathbb{X}} \sum_{i=1}^{N} \mathbf{s}_{i} \log(D^\tau(F(\mathbf{x}_{i}))) \\
+ (1 - \mathbf{s}_{i})\log(1 - D^\tau(F(\mathbf{x}_{i}))).
\end{multline}
Notably, only the parameters of $D^\tau$ are optimized, while the parameters of $F$ are kept fixed. The details of the RDAL method are fully outlined in Algorithm~\ref{alg:adv}. In this study, the capability of RDAL algorithm is enhanced by augmenting a masking U-net architecture prior to the feature extractor $F$, as outlined in \cite{luong2023representation}.
% As it is indicated by the Equation~\ref{eq:verify_loss}, only the parameters of $D_\tau$ are optimized, and the parameters of $F$ are kept fixed. RDAL is fully demonstrated in Algorithm~\ref{alg:adv}.
% After the optimization of $D_\tau$, it will be used to replace $D$ in the adversarial learning process. RDAL is fully demonstrated in Algorithm~\ref{alg:adv}.

\begin{algorithm}[tb!]
   \caption{Robust discriminative adversarial learning (RDAL)}
   \label{alg:adv}
\begin{algorithmic}
   \STATE {\bfseries Require:} Labelled data $\mathbb{X}$, trainable networks parameters: $F_\theta$, $C_\theta$, $D_\theta$, $D^\tau_\theta$
   \STATE {\bfseries Require:} Learning rate $\mu$, batch size $B$, training $D^\tau_\theta$ every  $\tau$ epochs
   \STATE {\bfseries Require:} GRL multiplier $\lambda(m)$ at epoch $m$
   % \REPEAT
   \WHILE{{\bfseries NOT} converged}
   % \FOR{$1 \cdot B$}
   \STATE Sample a mini-batch $\{(\mathbf{x}_{i}, \mathbf{y}_{i}, \mathbf{s}_{i})\}_{i=1}^{B}$  
   \STATE $\mathcal{L}_{cls} = \text{CE}(\{(C_\theta(F_\theta(\mathbf{x}_i)), \mathbf{y}_i)\}_{i=1}^{B})$
   \STATE $\mathcal{L}_{adv} = \text{BCE}(\{(D_\theta(F_\theta(\mathbf{x}_i)), \mathbf{s}_i)\}_{i=1}^{B})$
   \STATE $\theta_C \longleftarrow \theta_C - \mu \left(\frac{\partial \mathcal{L}_{cls}}{\partial \theta_C}\right)$, \STATE
   $\theta_D \longleftarrow \theta_D - \mu \left(\frac{\partial \mathcal{L}_{adv}}{\partial  \theta_D}\right)$, \STATE
   $\theta_F \longleftarrow \theta_F - \mu \left(\frac{\partial \mathcal{L}_{cls}}{\partial  \theta_F} - \lambda \frac{\partial \mathcal{L}_{adv}}{\partial \theta_F}\right)$
   % \ENDFOR
   \IF{$m \bmod \tau = 0$}
   \STATE Initialize $D^\tau_\theta$
   \WHILE{{\bfseries NOT} converged}
   \STATE Sample a mini-batch $\{(\mathbf{x}_{i}, \mathbf{s}_{i})\}_{i=1}^{B}$ 
   \STATE $\mathcal{L}_{sp} = \text{BCE}(\{(D^\tau_\theta(F_\theta(\mathbf{x}_i)), \mathbf{s}_i)\}_{i=1}^{B})$
   \STATE $\theta_{D^\tau} \longleftarrow \theta_{D^\tau} - \mu \left(\frac{\partial \mathcal{L}_{sp}}{\partial  \theta_{D^\tau}}\right)$
   \ENDWHILE
   \STATE $\theta_D \longleftarrow \theta_{D^\tau}$
   \ENDIF
   \ENDWHILE
\end{algorithmic}
\end{algorithm}

\section{Evaluation}
\label{sec:eval}

% In this section, we evaluate the performance of our method for obfuscating the speech presence in the latent feature space while recognizing targeted sound events as the main utility task. We assess the effectiveness of preserving privacy in the audio recordings by performing speech classification in one-second segments of audio. Likewise, the effectiveness of the utility task is the classification of sound events in these one-second segments. 
In this section, we evaluate the performance of our method in obfuscating speech information in the latent feature space while simultaneously performing sound event recognition for targeted classes as the primary utility task\textsuperscript{1}\def\thefootnote{1}\footnotetext{\url{https://github.com/lndip/RDAL}}. We measure the effectiveness of preserving privacy in audio recordings by conducting speech presence classification and gender classification in one-second audio segments. Additionally, we assess the performance on the utility task, which involves classifying sound events within these one-second segments. In our evaluation the segments are isolated from each other and not processed as a part of continuous audio, but we term the tasks as speech activity detection (SAD), gender detection (GD), and sound event detection (SED), to indicate that in a realistic application scenario we would be processing a continuous stream of segments, and segment-wise classification would lead to detecting the temporal activities of classes.

We compare the performance of RDAL against \textit{baseline} and naive adversarial methods. The \textit{baseline} is defined as a sound event classification system where no privacy measures are taken, meaning $F$ and $C$ are optimized jointly using Equation~\ref{eq:l_cls} in a supervised manner without the adversarial branch. In addition, the naive adversarial method does not include $D^\tau$ and therefore $F$, $C$, $D$ are optimized only using Equations~\ref{eq:l_cls}, \ref{eq:L_adv}, and \ref{eq:param_update}. We refer to this method in the rest of the paper as \textit{NaiveAdv}. In order to augment the capabilities of the RDAL algorithm and further improve its performance, we incorporate the masking network, as detailed in our preliminary study \cite{luong2023representation}. This enhanced variant is denoted as RDAL+M in this paper. By integrating a U-Net architecture of DNNs prior to the feature extractor $F$, RDAL+M aims to separate the speech component from the magnitude spectrogram of each data sample and reconstruct a non-speech version of that sample. The masking network is pre-trained and remains unchanged throughout the adversarial training process. More details of the masking approach are provided in \cite{luong2023representation}. 
% The dataset and code for our method will be made publicly available upon publication.
\subsection{Dataset}

In order to address the problem formulation outlined in Section~\ref{method}, it is necessary to use audio recordings which include both targeted sound events and speech. To achieve this, we create a simulated dataset using real-world audio recordings to generate one-second mixtures containing speech and other sound events. 

% We obtain the sound event data from the FSD50K dataset \citep{fonseca22}. From the 144 leaf nodes of the FSD50K dataset, we select 12 specific sound event classes that could potentially be used in an acoustic monitoring application and are listed in Table~\ref{tab:dataset}. For each sample from the target sound event classes, we extract the two most energetic one-second segments to have enough audio segments of each sound event classes. These segments are then normalized by subtracting their mean and dividing by their standard deviation. If a recording is shorter than one second, it is zero-padded at the end after normalization. The processed samples are divided into the \textit{development} and \textit{test} splits following the ``development'' and ``evaluation'' splits of the FSD50K dataset.
We collect the sound event data from the FSD50K dataset \citep{fonseca22}. Among the 144 leaf nodes in the FSD50K dataset, we select 12 specific sound event classes that are potentially applicable in acoustic monitoring applications. These selected classes are listed in Table~\ref{tab:dataset}. For each sample belonging to the target sound event classes, we extract the two most energetic one-second segments to ensure an adequate number of audio segments for each sound event class. These segments are then normalized by subtracting their mean and dividing by their standard deviation. If a recording is shorter than one second, we pad it with zeros at the end after normalization. The processed samples are divided into the \textit{development} and \textit{test} splits, following the ``development'' and ``evaluation'' splits defined in the FSD50K dataset.

% We use LibriSpeech corpus \citep{panayotov15} as the source for providing the speech contents of our dataset. We resample the recordings from LibriSpeech corpus from 16 kHz to 44.1 kHz to match the sampling frequency of the samples in FSD50K dataset. We follow the same process as discussed for the samples in FSD50K to extract the most energetic one-second speech segment from each audio recording. Then, these extracted segments of speech recordings from LibriSpeech \textit{train-clean-100} and \textit{dev-clean} sets are first attenuated by 5 dB, then utilized to be mixed with the selected one-second segments of sound events from FSD50K \textit{development} and \textit{evaluation} sets respectively.
% We use the LibriSpeech corpus \citep{panayotov15} as the source of speech content for our dataset. To match the sampling frequency of the FSD50K dataset, we resample the recordings of LibriSpeech corpus from 16 kHz to 44.1 kHz. Following the same process as described for the FSD50K samples, we extract the most energetic one-second speech segment from each audio recording and normalize the extracted segments. These segments which are selected from the LibriSpeech ``train-clean-100'' and ``dev-clean'' sets, are initially attenuated by 5 dB and mixed with the processed one-second segments of sound events from the \textit{development} and \textit{test} sets respectively.
The speech content for our dataset is sourced from the LibriSpeech corpus \citep{panayotov15}. To match the sampling frequency of the FSD50K dataset, we resample the recordings from 16 kHz to 44.1 kHz. Similar to the procedure used for the FSD50K samples, we extract the most energetic one-second speech segment from each audio recording and normalize these segments. The selected segments are obtained from the LibriSpeech ``train-clean-100'' and ``dev-clean'' sets. These segments are initially attenuated by 5 dB, then mixed with the processed one-second segments of sound events from the \textit{development} and \textit{test} sets respectively.

% The selection of speech recordings from LibriSpeech corpus is such that we ensure to have equal number of male and female speakers per each class of targeted sound event. In addition, the number of extracted one-second segments of speech recordings from LibriSpeech is half of the number sound event segments in order to create mixtures in two classes of speech and non-speech with balanced number of samples.
% The speech segments in the \textit{development} set are extracted from 126 male and 125 female speakers, and those in \textit{test} set are chosen from 20 male and female speakers. 
% The selection of speech recordings from LibriSpeech corpus ensures an equal number of male and female speakers per each class of targeted sound event. F
% we extract half the number of one-second speech segments from LibriSpeech compared to the number of sound event segments. 
% This approach allows us to create mixtures with a balanced number of samples in two classes: speech and non-speech.
% The \textit{development} set contains speech content from 126 male and 125 female speakers, while the \textit{test} set includes 20 speakers from male and female separately. When selecting speech recordings from the LibriSpeech corpus, we ensure an equal representation of male and female speakers for each targeted sound event class. Furthermore, The number of extracted one-second speech segments from LibriSpeech is half of the number of sound event segments. This approach allows us to create mixtures with a balanced number of samples for speech and non-speech classes.
The \textit{development} set comprises speech content from 126 male and 125 female speakers, while the \textit{test} set consists of 20 speakers from both genders. In selecting speech recordings from the LibriSpeech corpus, we ensure an equal representation of male and female speakers across each targeted sound event class. Furthermore, the number of extracted one-second speech segments from LibriSpeech is half of the number of sound event segments. This approach allows us to create mixtures with a balanced number of samples for both the speech and non-speech classes.
\begin{table}[t]
    \caption{Number of one-second sound event samples in each split of our dataset.}
    \begin{center}
    \begin{tabular}{|c|c|c|c|}
    \hline
    \textbf{Sound events} & \textbf{Train} & \textbf{Validation} & \textbf{Test} \\
    \hline
    \text{Dog barking} & 608 & 66 & 96 \\
    \text{Glass breaking} & 480 & 52 & 62 \\
    \text{Gun shot} & 469 & 52 & 179 \\
    \text{Cough} & 384 & 42 & 132 \\
    \text{Slam} & 383 & 42 & 93 \\
    \text{Applause} & 425 & 46 & 62 \\
    \text{Dishes, pot, and pan} & 298 & 36 & 73 \\
    \text{Toilet flush} & 324 & 36 & 52 \\
    \text{Cat meowing} & 208 & 22 & 94 \\
    \text{Doorbell} & 174 & 18 & 49 \\
    \text{Crying} & 171 & 18 & 40 \\
    \text{Drill} & 268 & 28 & 61 \\
    % \hline
    % \multicolumn{2}{|c|}{\text{Speech duration}} & 2094 & 227 & 494 \\
    % \hline
    % \multicolumn{2}{|c|}{\text{Male speakers}} &\multicolumn{2}{c|}{126} & 20\\
    % \multicolumn{2}{|c|}{\text{Female speakers}} &\multicolumn{2}{c|}{125} & 20\\
    \hline
    \end{tabular}
    \label{tab:dataset}
    \end{center}
\end{table}
% After creating our mixtures, we obtain two disjoint sets of data which we call \textit{development} and \textit{test} sets. In order to have separate data for model selection purposes during training, the \textit{development} set is randomly split into \textit{train} and \textit{validation} with the ratio of 9:1. The division of data also ensures that half of the mixtures within each split, belonging to each sound event class, contain speech. However, no further constraint about speakers are applied, thus the speakers in the \textit{train} and \textit{validation} splits are not exclusive, and the ratio of male and female speakers are not necessarily balanced. Table~\ref{tab:dataset} shows the number of sound event samples in each split of our dataset, and the number of unique male and female speakers in the \textit{development} and \textit{test} splits.
% Once the mixtures are created, we obtain two distinct sets of data referred to as the \textit{development} and \textit{test} sets. 

To facilitate model selection during training, the \textit{development} set is further divided randomly into the \textit{train} and \textit{validation} sets in a 9:1 ratio. This division ensures that half of the mixtures within each split, across sound event classes, contain speech. Specifically, the number of samples containing speech in \textit{train}, \textit{validation}, and \textit{test} are 2094, 227, and 494, respectively. However, there are no specific constraints on speaker allocation. Therefore, the speakers in the \textit{train} and \textit{validation} sets may overlap, and the ratio of male to female speakers may not be balanced within the two splits. 
% Table~\ref{tab:dataset} provides an overview of the number of sound event samples for each split in our dataset, as well as the number of unique male and female speakers in the \textit{development} and \textit{test} sets.

% Log-mel spectrograms of audio mixtures are used as our low-level features $\mathbf{x}$. The parameters of this transformation are based on \citet{kong20}; the values are scaled based on the sampling frequencies of the our audio recordings (44.1 kHz vs 32 kHz in \citep{kong20}). For the short-time Fourier transform (STFT) calculation, we use a Hamming window with size 1411, and hop length of 441. 64 mel filter banks are used to calculate the log-mel spectrograms. For RDAL+M method, the magnitude of STFTs is employed as the input to the masking network; the log-mel spectrogram is calculated on the masked spectrogram before being passed through $F$.
We utilize log-mel spectrograms as the low-level features, denoted as $\mathbf{x}$, for the audio mixtures. The parameters for this transformation are derived from the work of \citet{kong20} but are adjusted to account for the sampling frequencies of our audio recordings (44.1 kHz instead of 32 kHz as used in \citep{kong20}). In computing the short-time Fourier transform (STFT), we apply a Hamming window of size 1411 with a hop length of 441. To obtain the log-mel spectrograms, we employ 64 mel filter banks. In the RDAL+M method, the magnitude of the STFTs serves as the input to the masking network, while the log-mel spectrogram is calculated on the masked spectrogram prior to being passed through the feature extractor $F$.

\begin{figure}[t]
    \begin{center}
    \includegraphics[width=0.7\columnwidth]{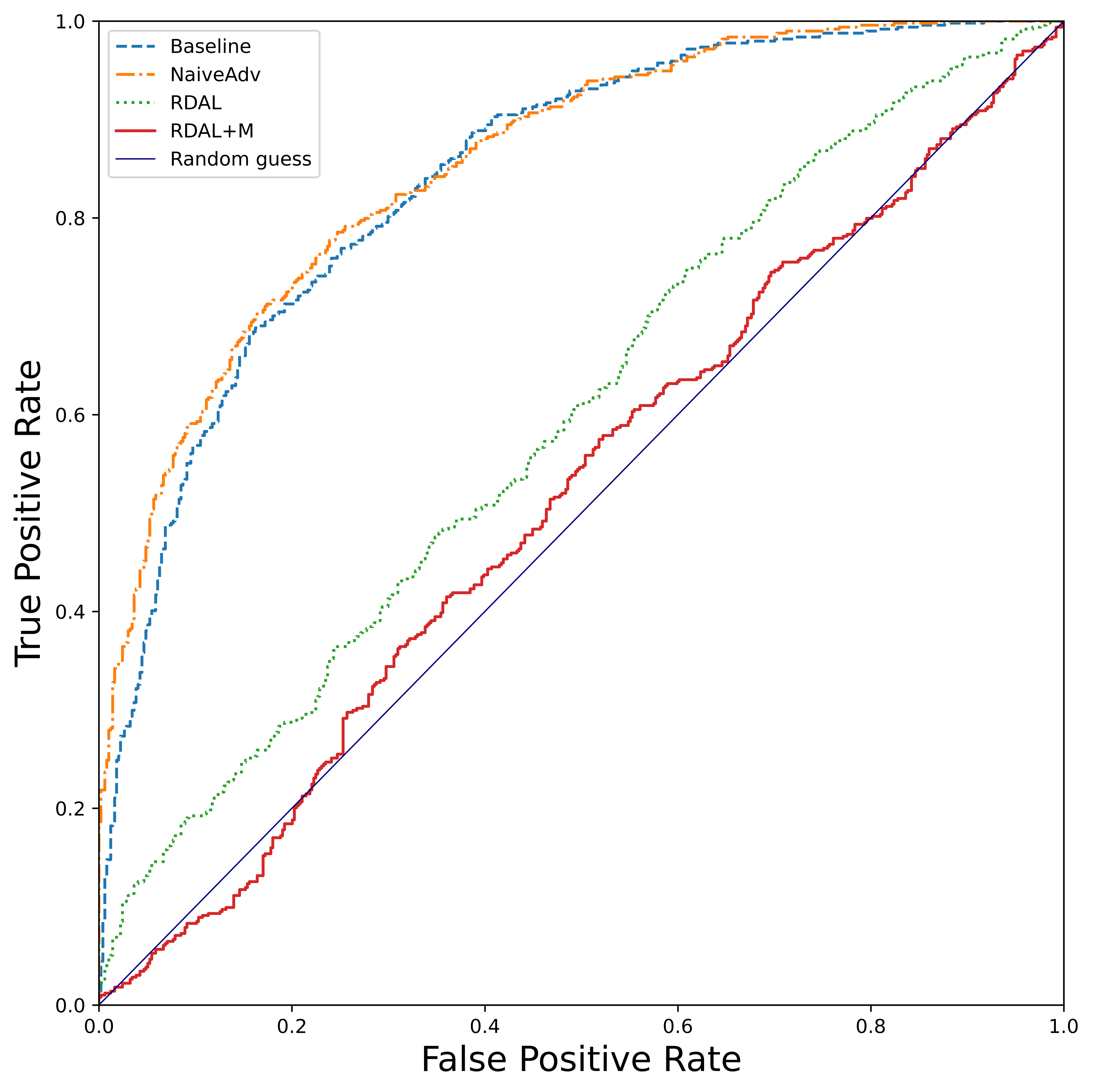}
    \caption{ROC curves for each method are displayed, showcasing the privacy preservation results on the SAD task as outlined in Table~\ref{results-table}.}
    \label{roc_curve}
    \end{center}
\end{figure}

\subsection{Network architecture}
% We use ``CNN6'' architecture of DNNs in \citet{kong20} as our feature extractor $F$ with minimal changes to match the size of our data. Therefore, our adapted architecture includes 4 convolutional blocks; each block includes a 2D convolutional layer with kernel size $3 \times 3$, ReLU activation function, and batch normalization. The number of convolutional filters in the blocks are 64, 128, 256 and 512, respectively. In addition, the convolutional blocks, apart from the last one, have max pooling with the kernel size of $2 \times 2$. 
% The last convolutional block uses max global pooling that converts our 2D features into 1D vector representations. Finally, a linear layer is used to outputs the latent features $\bf z$ as a vector with 64 elements. The sound event classifier $C$ is a single linear layer with softmax activation. The architecture of the speech classifier $D$ is composed of 4 linear layers. The first three layers consists of 48, 32, and 16 output dimensionality respectively and each layer is followed by a LeakyReLU activation function. The output layer of $D$, however, includes sigmoid as the activation function.
For our feature extractor $F$, we utilize the ``CNN6'' architecture from \citet{kong20} as a basis, making minimal adjustments to accommodate our data size. Our adapted architecture consists of 4 convolutional blocks, each containing a 2D convolutional layer with a kernel size of $3 \times 3$, ReLU activation, and batch normalization. The number of convolutional filters in these blocks is 64, 128, 256, and 512, respectively. Except for the last block, all blocks incorporate max pooling with a kernel size of $2 \times 2$. The final convolutional block employs max global pooling to transform the 2D features into a 1D vector representation. A linear layer is then utilized to generate the latent features $\mathbf{z}$ as a 64-element vector. The sound event classifier $C$ consists of a single linear layer with softmax activation. As for the speech classifiers $D$ and $D^\tau$, it comprises 4 linear layers, with output dimensionality of 48, 32, and 16 in the first three layers, respectively, each followed by a LeakyReLU activation function. The output layer employs sigmoid as the activation function. For RDAL+M, we employ the exact architecture of masking network described in \cite{luong2023representation}.
\begin{table*}[h!]
    \centering
    \caption{Results of \textit{baseline}, \textit{NaiveAdv}, RDAL, RDAL+M.}
    \begin{tabular}{c|c|cc|cc}
    \toprule
    Methods & SED\textsubscript{accuracy} & SAD\textsubscript{accuracy} & SAD\textsubscript{AUC score} & GD\textsubscript{accuracy} & GD\textsubscript{AUC score} \\
    \midrule
    Baseline & $0.75\pm0.01$ & $0.76\pm0.01$ & $0.84\pm0.01$ & $0.63\pm0.02$ & $0.68\pm0.03$\\
    NaiveAdv & $0.75\pm0.01$ & $0.75\pm0.02$ & $0.83\pm0.02$ & $0.59\pm0.02$ & $0.63\pm0.03$\\
    \midrule
    RDAL ($\tau$=$50$) & $0.75\pm0.01$ & $0.62\pm0.03$ & $0.68\pm0.04$ & $0.59\pm0.03$ & $0.63\pm0.03$\\
    RDAL+M ($\tau$=$70$) & $\mathbf{0.77\pm0.01}$ & $\mathbf{0.52\pm0.02}$& $\mathbf{0.52\pm0.02}$& $\mathbf{0.50\pm0.02}$ & $\mathbf{0.50\pm0.02}$\\
    \bottomrule
    \end{tabular}
    \label{results-table}
\end{table*}
\subsection{Training}
% We set $\lambda = -1$ 
% We train the entire networks in a supervised way for the first 30 epochs ($\lambda$ = 0). This is due to the known stability issues of adversarial training specially at the starting iterations \cite{acuna2022domain}, and to ensure that the speech classifier $D$ is adequately trained to distinguish between speech and non-speech classes from the very beginning of the adversarial training. Following the first 30 epochs, adversarial training is initiated by gradually increasing the value of $\lambda$, starting from 0. Once the maximum value of 1 is reached during the training process, $\lambda$ is fixed. This is achieved using the following schedule adapted from \citet{pmlr-v37-ganin15}:
Initially, we train the entire networks in a supervised manner for the first 30 epochs ($\lambda$ = 0). This helps address stability issues associated with adversarial training during the initial iterations \cite{acuna2022domain} and ensures proper training of the speech classifier $D$ to recognize the speech presence before the start of adversarial training. Following this, we gradually increase the value of $\lambda$ to initiate adversarial training. Once the maximum value of 1 is reached, $\lambda$ remains fixed. The schedule for increasing $\lambda$ is adapted from \citep{pmlr-v37-ganin15} and is defined as: 
\begin{equation}
\lambda_{\beta} = \frac{2}{1+\exp(-\gamma.\beta)}-1
\end{equation}
where $\gamma$ is set to 100 in all experiments, and $\beta \in [0, 1]$ represents the progress of the training process. After the first 30 epochs, $\beta$ starts at 0 and increases with a step size determined by dividing the range from 0 to 1 into equal intervals over the course of the epochs, with a maximum number of 5000 epochs. The initial epoch for starting adversarial training has not been optimized.
% increases incrementally after each epoch to reach its maximum at 1. 
% The initial epoch for the start of adversarial training has not been optimized. 
% The optimized value of $\tau$, which indicates the number of epochs to run the adversarial training before training a new $D^\tau$, was searched over the values in \{10, 20, 30, 50, 70, 100\}. We select the best value using the validation set. The models are trained on batch sizes of 64 where half of the samples contain speech. Stochastic gradient descent (SGD) with learning rate of 0.01 and momentum of 0.9 is used to optimize the parameters of our networks. We use the best $\mathcal{L}_{sp}$ value for the validation set as the criterion for early stopping such that if it does not increase after 10 repetitions, the training process will stop.
The optimal value of $\tau$, indicating the number of epochs for adversarial training before training a new $D^\tau$, is selected from the values in the set of \{10, 20, 30, 50, 70, 100\} using the validation set. The models are trained with a batch size of 64, with half of the samples containing speech. Stochastic gradient descent (SGD) with a learning rate of 0.01 and momentum of 0.9 is employed for optimizing the networks parameters. Early stopping is determined by monitoring the best $\mathcal{L}_{sp}$ value on the validation set, and if there is no improvement after 10 repetitions, the training process is halted.

\subsection{Results}
% Table~\ref{results-table} presents the accuracy for SED as well as the accuracy and recall for SAD over the test set of our dataset. After finishing the training process, in order to be able to verify the performance of \textit{baseline}, \textit{NaiveAdv}, RDAL, and RDAL+M in preserving the speech privacy, we train a new classifier (we call this classifier attacker model) over the latent representations $\bf z$ to distinguish between speech and non-speech samples. This simulates an attack scenario where an intruder tries to recognize speech activities within one-second segments. All presented values in Table~\ref{results-table} are calculated by averaging over 10 separate runs. 
% Table~\ref{results-table} displays the accuracy of SED as well as the accuracy and AUC scores of SAD and GD tasks on the test set. Once the training is completed, we train a new classifier (referred to as the attacker model) on the latent representations $\mathbf{z}$ to assess the speech privacy preservation performance of the \textit{baseline}, \textit{NaiveAdv}, RDAL, and RDAL+M methods. This classifier dependent on the task aims to identify speech presence or gender, simulating a scenario where an intruder attempts to recognize speech activities in one-second segments using latent features. All values in Table~\ref{results-table} are averaged over 10 separate runs.
Table~\ref{results-table} presents the accuracy results for SED, as well as the accuracy and AUC scores for SAD and GD tasks on the test set. After completing the training phase, we proceed to train a new classifier, referred to as the attacker model, using the latent representations $\mathbf{z}$. This classifier is specifically designed for the task of identifying speech presence or gender. It simulates a scenario where an unauthorized individual attempts to recognize speech activities in one-second segments using latent features. The evaluation in Table~\ref{results-table} is based on the average values obtained from 10 separate runs.

The results presented in Table~\ref{results-table} yield several noteworthy observations. Firstly, the comparable SED accuracy scores across the three adversarial methods indicate that optimizing the adversarial branch to eliminate speech activity, while simultaneously training a supervised SED system, does not hinder the optimization of the SED task. Furthermore, RDAL demonstrates a notable improvement in privacy preservation compared to the \textit{baseline}, as evidenced by the SAD and GD metrics. Additionally, RDAL+M further enhances RDAL's privacy preservation performance, achieving a reduced accuracy and AUC score for both SAD and GD tasks, approaching the level of random guess scores in binary classification tasks. Thirdly, the SAD accuracy of \textit{NaiveAdv} suggests that it does not offer enhanced privacy preservation compared to the \textit{baseline} when evaluated against the attacker model. Therefore, \textit{NaiveAdv} does not truly provide privacy-preserved features. Lastly, the lower GD performance across all methods indicates the inherent difficulty of this task compared to SAD. Given the absence of specific gender information during training, the scores for this task generally fall below those of SAD, making it relatively easier to obfuscate in the context of privacy preservation.
\begin{figure}[t!]
    \centering
    \includegraphics[width=0.45\columnwidth]{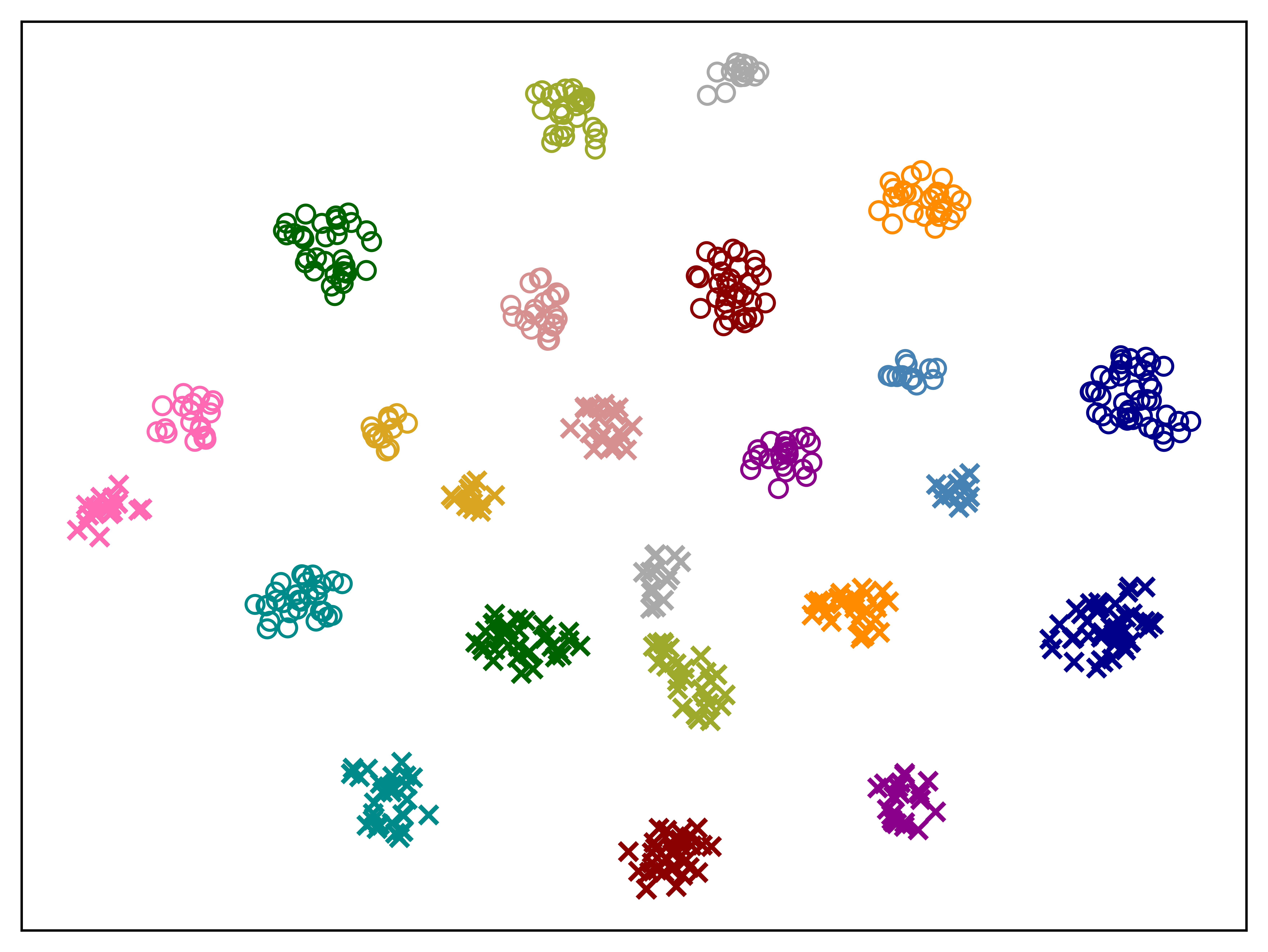}
    \includegraphics[width=0.45\columnwidth]{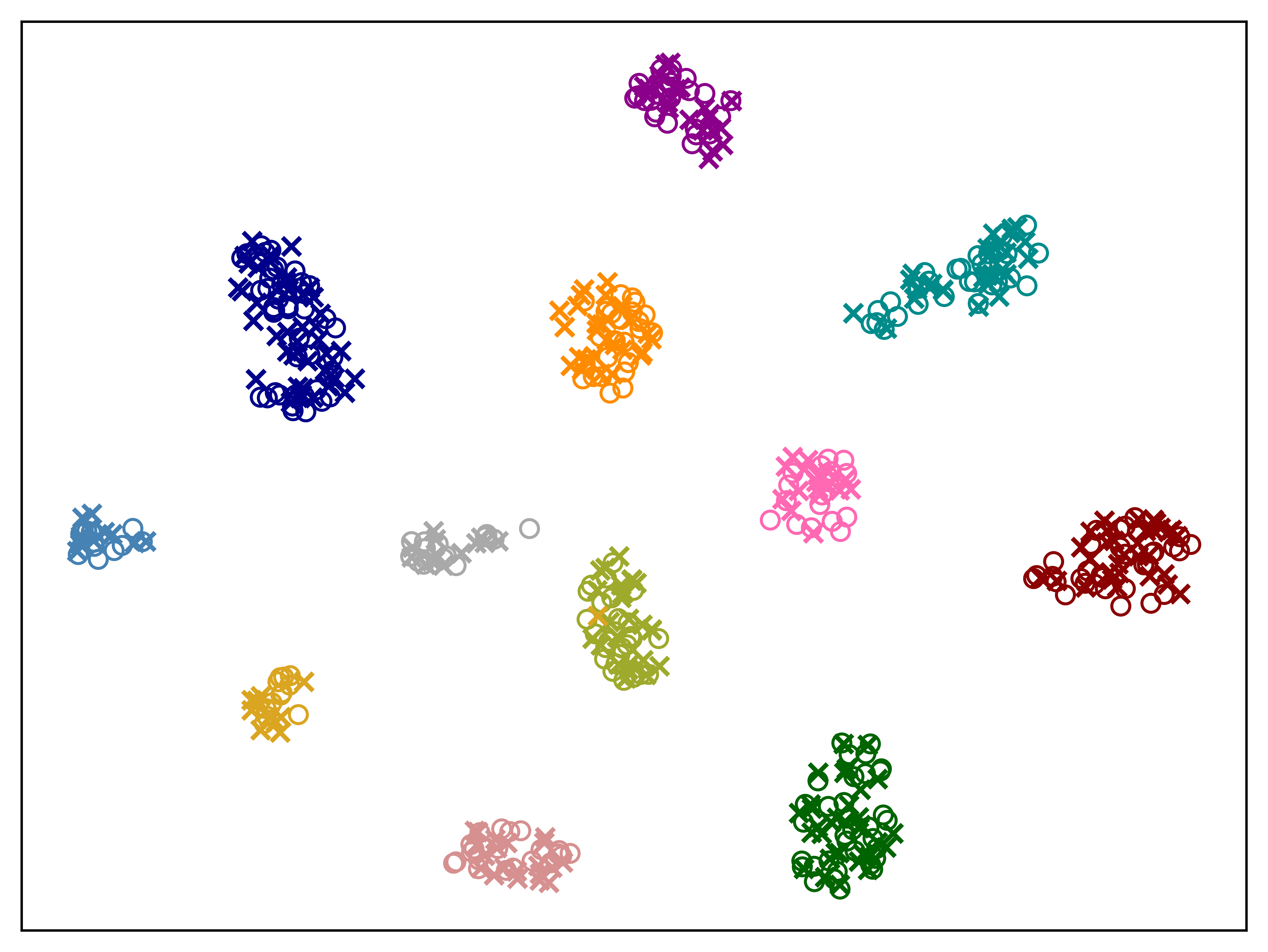}
    \caption{Comparison of latent features obtained by RDAL's $F$ (right) and supervised training of $F$ for sound events and speech (left). Sound events are color-coded with 12 different colors, while speech and non-speech samples are marked with ``o'' and ``x'' respectively.}
    \label{tsne-analysis-plots}
\end{figure}
\subsection{Discussion}
To ensure a fair comparison among the methods presented in Table~\ref{results-table}, we ensure the adoption of an identical architecture for DNNs models across all methods. In addition, the same scheduling of $\lambda$ values is used for the \textit{NaiveAdv}, RDAL, and RDAL+M methods. 
% The performance of \textit{RDAL} is reported for different $\tau$ values in Figure~\ref{P-performance-plot}. 
% The notable improvement in preserving the privacy of RDAL over \textit{baseline} method proves the main claim of the paper. We also clarify that a naive implementation of an adversarial training method will not be effective at all due to the existing limitation in optimizing the feature extractor for aligning the two distributions of speech and non-speech classes, hence the incapability of \textit{NaiveAdv} in preserving privacy in our problem setup. \citet{srivastava2019} previously showed that privacy-preserving performance presented by \textit{NaiveAdv} compared to the achieved results by \textit{baseline}, which does not use any privacy preserving components in its entirety, can be even worse in an open-set classification. 
% They performed speaker verification in an open-set setup where the aim is to conceal the speaker's identity for unknown speakers, and the adversarial training improved the performance in speaker verification. Similarly, our test data is composed of audio recordings of speech contents that have not been seen in our train and validation sets. 

The significant privacy preservation improvement achieved by RDAL in comparison to the \textit{baseline} method substantiates the main claim of this paper. Furthermore, we emphasize that a naive implementation of adversarial training proves ineffective due to inherent limitations in optimizing the feature extractor to align speech and non-speech distributions. Consequently, the \textit{NaiveAdv} method fails to effectively preserve privacy within our problem setup. Previous work by \citet{srivastava2019} has demonstrated that the privacy-preserving performance of \textit{NaiveAdv} can even be worse than that of the \textit{baseline} method, which lacks any privacy-preserving components, particularly in open-set classifications.

\begin{figure*}[h!]
    \begin{center}
    \includegraphics[width=0.32\textwidth]{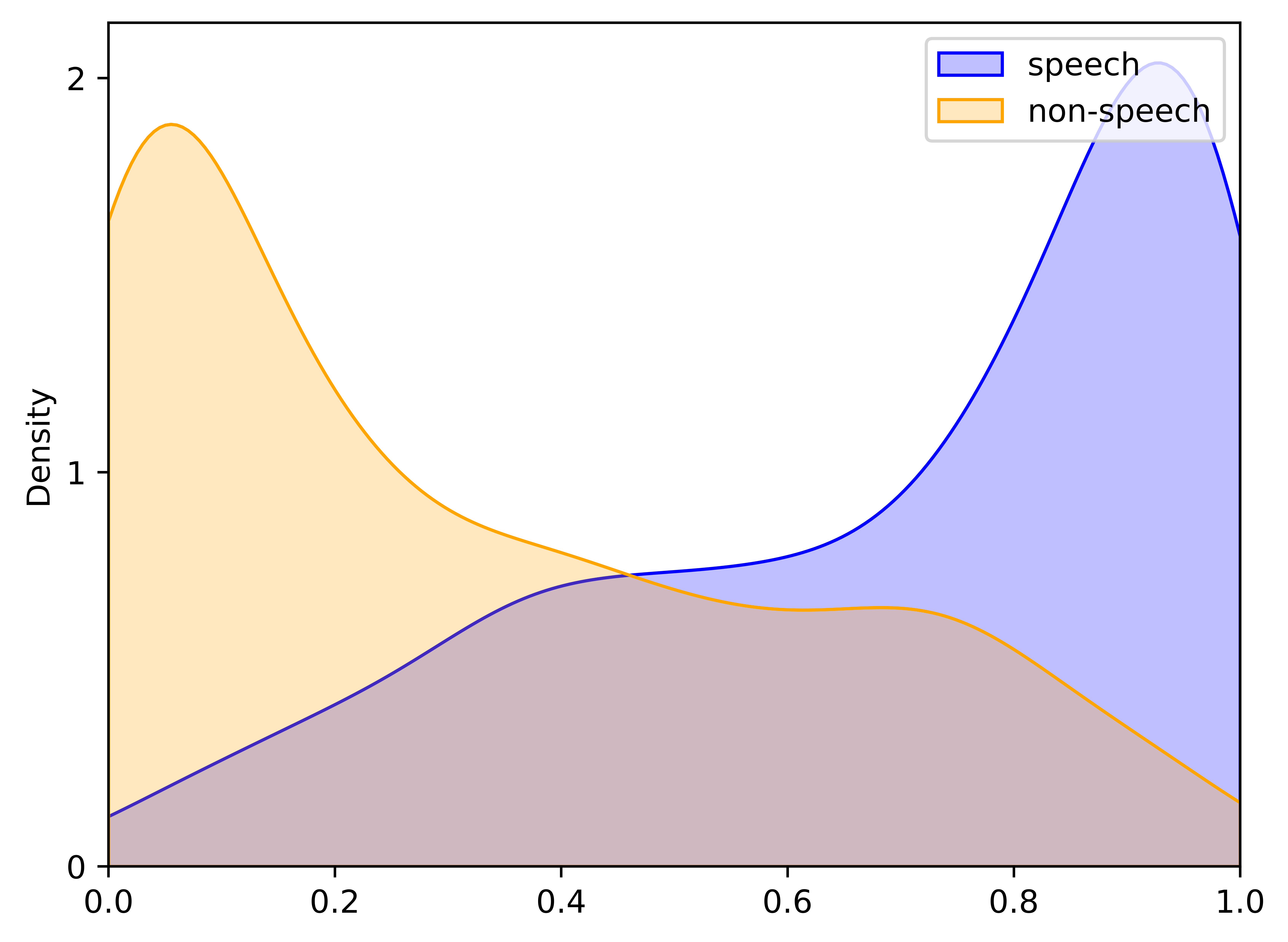}
    \includegraphics[width=0.32\textwidth]{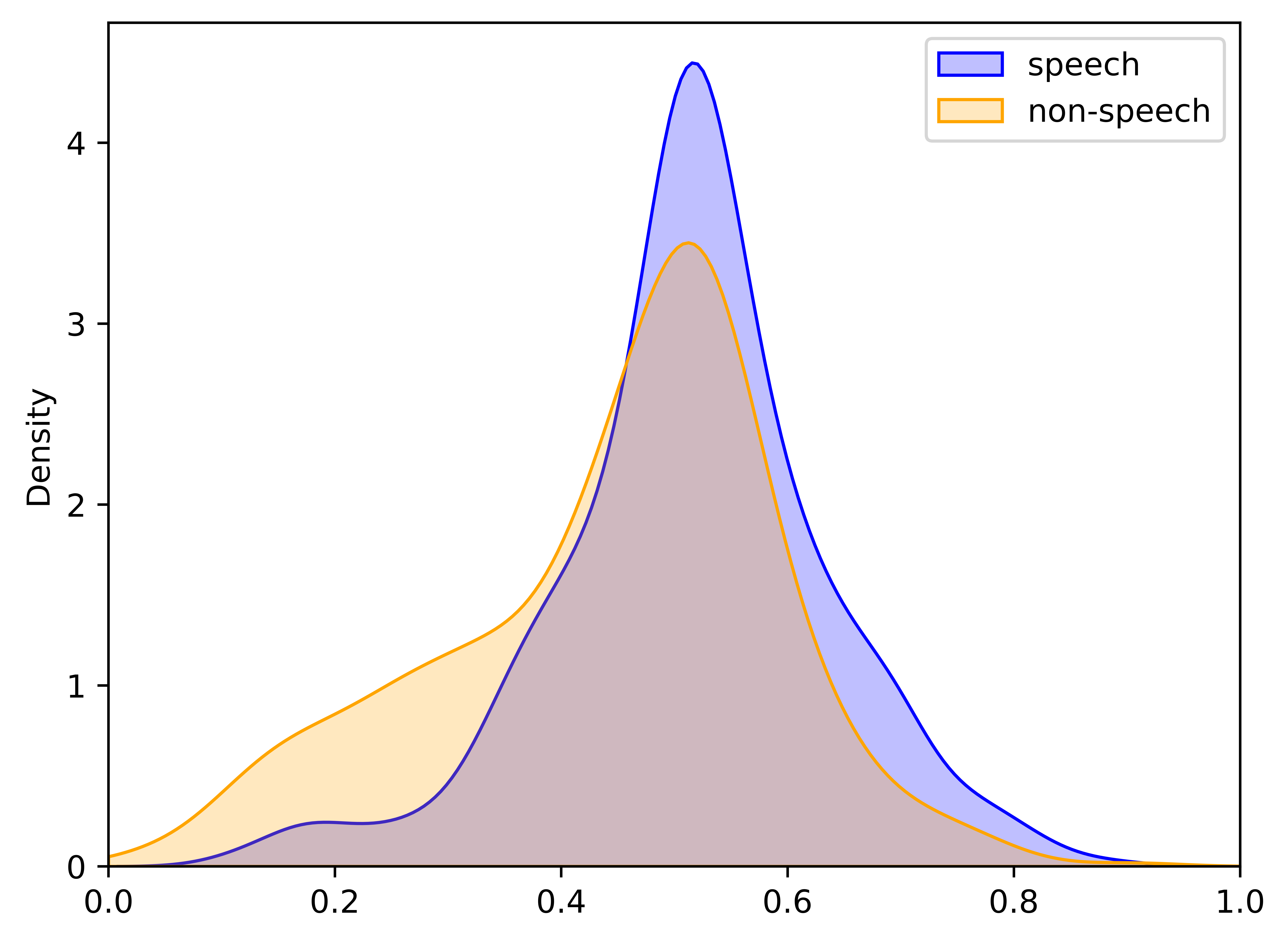}
    \includegraphics[width=0.32\textwidth]{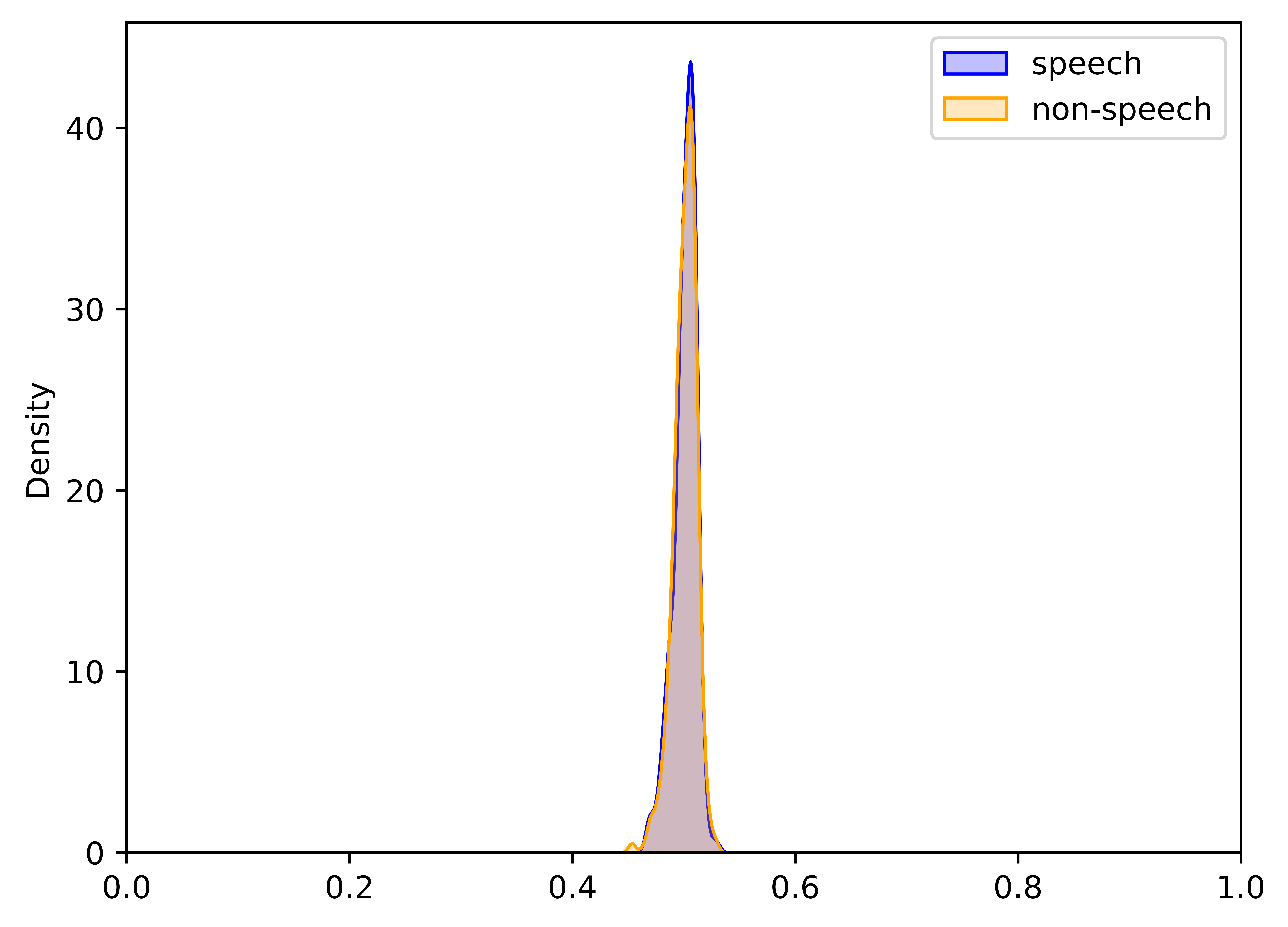}
    % \caption{Estimated density curves using Gaussian kernel to represent continuous probability densities of the predicted probabilities for the test data using \textit{baseline} (left figure) and \textit{RDAL+M} (right figure) methods.}
    \caption{Density curves using Gaussian kernel to represent predicted probability densities from the attacker model on the test data using the latent features of \textit{baseline} (left),  RDAL (middle), and RDAL+M (right) methods.}
    \label{disc-prob-plot}
    \end{center}
\end{figure*}

% Additionally, Figure~\ref{tsne-analysis-plots} illustrates the distributions of latent features $\bf z$ in 2D using t-SNE analysis \citep{vandermaaten08}. We compare the feature distributions of RDAL to a system acting as a lower bound for privacy preservation where the supervised information of targeted sound events and speech presence are provided into the system without any regards for preserving the privacy. The comparison demonstrates that the feature distributions of speech and non-speech classes are properly aligned in each targeted sound event class, further corroborating the enhanced privacy-preserving performance of RDAL. 
Figure~\ref{tsne-analysis-plots} showcases the 2D distributions of latent features $\bf z$ using t-SNE analysis \citep{vandermaaten08}. We compare RDAL's feature distributions with a privacy-preserving lower bound system that incorporates supervised information of targeted sound events and speech presence without considering privacy preservation measures. The comparison reveals proper alignment of speech and non-speech class distributions within each targeted sound event class, providing further evidence of RDAL's improved privacy preservation performance.

Figure~\ref{disc-prob-plot} visualizes the kernel density estimates representing the predicted probabilities for the speech and non-speech classes. These probabilities are computed using the test set after training the attacker model. In the \textit{baseline} method (left figure), the density curves clearly indicate the attacker's confident predictions regarding the presence or absence of speech. This highlights the information embedded in the latent representations of SED systems, even when speech is not the target sound event. In contrast, the density curves of RDAL (middle figure) exhibit significant overlap between the speech and non-speech classes, indicating an increase in model uncertainty and the attacker's challenge in distinguishing between the two. This increased uncertainty in the attacker model's predictions is a result of aligning speech and non-speech latent features through the optimized feature extractor from the RDAL method. Furthermore, RDAL+M (right figure) improves upon the results achieved by RDAL, and demonstrates a precise alignment between the density curves of predicted probabilities for the speech and non-speech classes. Given that the attacker model performs binary classification, uncertainty in its predictions can be quantified using the binary entropy function. Maximum uncertainty arises when the attacker predicts a sample with a probability of 0.5. As evidenced by the Figure~\ref{disc-prob-plot}, a substantial portion of the density mass for both speech and non-speech classes is concentrated around this value.

%The uncertainty of speech presence predictions can be quantified, for example, using the binary entropy function for a 
%The uncerainty of This indicates maximum uncertainty in the attacker's predictions.
%In our case, the maximum entropy occurs when the speech classifier $D$ predicts a sample with a probability of $p = \frac{1}{2}$, indicating maximum uncertainty regarding the sample's class.

% Specifically, the increased alignment of latent features leads to greater uncertainty in the predictions of the speech classifier $D$. This uncertainty can be quantified using the entropy of a binary trial, given by:
% \begin{equation}
% H(\mathbb{P}) = -p\log(p) - (1-p)\log(1-p)
% \end{equation}
% where $\mathbb{P} = [p, 1-p]$ represents the probabilities of predicting a sample as speech or non-speech, respectively. In our case, the maximum entropy occurs when the speech classifier $D$ predicts a sample with a probability of $p = \frac{1}{2}$, indicating maximum uncertainty regarding the sample's class. Figure~\ref{disc-prob-plot} depicts this concept using the results of our method.

% This further clarifies the effectiveness of RDAL in increasing the privacy preservation.   

\section{Conclusion}
% SED systems are vulnerable to privacy breaches that can compromise the confidentiality of user information and related sensitive data. To address this issue, privacy-preserving algorithms are necessary to reduce the risk of revealing private information. This study formulates the problem of privacy preservation as detecting speech activity from the latent features of audio mixtures. We propose an adversarial training approach, called RDAL, to learn robust and speech-invariant latent features that are agnostic to speech presence and gender identity while preserving the information of targeted sound events for SED.
Privacy breaches pose a significant threat to the confidentiality of user information and sensitive data in SED systems. To mitigate this risk, it is crucial to employ privacy-preserving algorithms that safeguard against the disclosure of private information. This study addresses the issue by formulating privacy preservation as the detection of speech activity in the latent features of audio mixtures. We introduce RDAL, an adversarial training approach, which learns robust and speech-invariant latent features. RDAL ensures agnosticism towards speech presence and gender identity, while preserving the targeted sound event information for SED.

The proposed method utilizes two neural networks: a feature extractor and a speech classifier, in a minimax game to ensure the privacy preservation of audio mixtures. The feature extractor generates invariant latent features of speech-containing audio signals that are indistinguishable from those of non-speech ones, while the speech classifier tries to distinguish between them. We also address the limitations of this approach by introducing a new speech classifier periodically into the adversarial training process to enforce the feature extractor to consistently improve the performance for aligning the distributions of speech and non-speech samples during the adversarial training process.

The empirical results indicate that the proposed RDAL approach significantly improves the privacy performance of SED systems. By effectively preserving privacy in latent features of audio mixtures, this approach can help prevent potential privacy violations and ensure the confidentiality of sensitive information. Furthermore, we demonstrated that the performance of RDAL can be further improved through its integration with a source separation method.

\bibliographystyle{unsrtnat}
\bibliography{bibtex}

\end{document}